\documentstyle[12pt]{article}
\date{}
\begin{document}

\title{Nucleon Resonance Transition Couplings to Vector Mesons}

\author{D.O. Riska$^1$ and G.E. Brown$^2$}
\maketitle

\centerline{\it $^1$Department of Physics, 00014 University of Helsinki,
Finland}

\centerline{\it $^2$Department of Physics, State University of New York,
Stony Brook, N.Y. 11794, USA}

\begin{abstract}
Recent heavy ion experiments indicate modifications of the
$\rho-$meson mass {\it in medium}. In the CERES experiments
$\rho-$mesons are produced at $\sim$ normal nuclear matter
density, where hadrons are more appropriate constituents than
quarks. A collective "nuclear $\rho$", in which every
nucleon is excited to the $N(1520)$ state, with equal
amplitude, enters in this description. 
At the higher densities reached by
future experiments constituent quarks
become the appropriate variables. Here 
the $\rho$ and $\omega$ transition couplings to the
nucleon resonances up to 1700 MeV, including the $N(1520)$, 
are derived by means of the chiral
quark model. The relevant coupling constants are expressed in 
terms of the
corresponding vector coupling constants to nucleons. The quality of
the model relations is tested by a calculation of the
corresponding pion-resonance coupling constants, which are known from
the empirical pion decay widths.

\end{abstract}

\newpage

\section{Introduction}

The observed enhancement \cite{CERES,CERES2,CERES3,HELIOS} of low mass dilepton
production in relativistic nucleon-nucleon collisions has
stimulated considerable theoretical activity. The concensus
in the theoretical treatment is that the dileptons seen in the
CERES experiments \cite{CERES} to date originate with
densities close to the density $\rho_0$ of normal nuclear matter.
At these densities hadronic variables are more suitable than
quarks in the theoretical treatment. A rather convincing
description is that based on the coupling of nucleon isobar-hole
excitations was initiated by Peters et al. \cite{Peters}
and later systematized by Rapp and Wambach \cite{Rapp}. A
relation between this approach and the overall scaling
relations proposed by Brown and Rho \cite{BroRho}, formulated
in quark language and possibly applicable at higher densities
has been given  in ref.\cite{Gerry}. More recently the
empirical data have been used to fix the parameters in an
effective field theory approach \cite{Friman,Friman1}. In this
relatively model independent treatment a good description
of the dilepton excess seen by in the CERES experiments is
given.

It has been argued that at higher densities, in the region
between nuclear matter density, and the critical density
of chiral symmetry restoration, hadronic variables should
be replaced by quark variables \cite{Bub}. The
example of the Nambu-Jona-Lasinio model, in which the
constituent quark mass is the order parameter, suggests that
hadron masses go to zero in the limit of bare current
quark masses. In ref.\cite{Gerry} this is brought about by
replacing the $\rho-$meson mass $m_\rho$, which sets the scales of the
denominator by the Rapp-Wambach Lagrangian \cite{Rapp}, 
by the effective $\rho-$meson mass $m_\rho^*$. While this
result may be obtained on the basis of self-consistency, the
issue of the most appropriate treatment at densities above
that of normal nuclear matter remains largely open. The
chiral quark model, which implies relations between the
vector meson transition couplings to the nucleon resonances
and the vector meson couplings to nucleons, may impose 
useful constraints on the theoretical treatment.

The chiral quark model, in which constituent quarks couple directly to
mesons, is known to describe the properties of the ground state octet
and decuplet baryons fairly well
\cite{Gloza}. In particular it gives expressions for
the $N-\Delta(1232)$ transition couplings, which are good at the level of
$\sim 25\%$ accuracy, or better. Moreover, when augmented with a linear
confining interaction, and two-pion and vector meson
interactions between the quarks, it describes the whole
empirical baryon spectrum satisfactorily \cite{Glozb,GEB}.

We here use this model to calculate the $\rho$- and $\omega$-meson 
transition couplings to the nucleon resonances up to 1700 MeV. These
coupling constants cannot be determined directly from empirical decay
widths, as they lie below the thresholds for vector meson decay.
The quality of the model is tested by a calculation of the corresponding
pion-resonance transition coupling
constants in the impulse approximation, which are then 
compared to the values that
are determined from the pionic decay widths.
That the single quark coupling model for pion decays should provide
a fair overall description is indicated by the fact that it
implies that the ratios of the decay widths for
$\Xi^*\rightarrow \Xi\pi$, $\Sigma^*\rightarrow\Sigma\pi$ and
$\Delta\rightarrow N\pi$ should be 1:4:9, which compares well to
the empirical ratio 1:3.9:12.6 (that the number for the $\Delta$ 
exceeds 9 is due to the larger phase space volume
available).
For the excited $S-$ and the $P-$shell resonances the quark model
values, which are calculated here in the impulse approximation,
fall within factors 1.5 -- 2 of the empirically
extracted transition couplings. The situation for the
$D-$shell resonances is less satisfactory. 
Improved agreement requires
taking into account higher order contributions from
two-quark operators.  

The vector meson transition coupling constants to nucleon
resonances are 
defined in terms of Lagrangian densities for the transition
couplings, which involve generalized Rarita-Schwinger 
vector spinors. Comparison of the matrix elements of these
Lagrangians to the corresponding matrix elements in the
quark model makes it possible to express the transition
coupling constants in terms of the corresponding vector 
meson coupling constants to the
nucleons. The latter are determined - albeit within a liberal
uncertainty range - by fits to nucleon-nucleon
scattering data with phenomenological boson exchange interaction models.
These expressions involve $SU(2)$ Clebsch-Gordan coefficients as well as
orbital matrix elements of quark wave functions, which connect the
$P-$ and $D-$shell and the excited $S-$shell 
states to the ground states. The latter
depend on the Hamiltonian model for the 3-quark system. We here employ a
simple covariant harmonic oscillator model based on linear confining
interaction with a flavor-spin dependent hyperfine interaction, 
which describes the empirical spectrum very well \cite{Coe1}.

There is some freedom in the choice of the resonance transition
coupling Lagrangians. This freedom is constrained by comparison
with the corresponding quark model expressions, especially
because of the orthogonality of the resonance and nucleon
wave functions in the quark model. The matching of 
matrix elements of covariant Rarita-Schwinger type 
Lagrangians \cite{Cheng} and quark model matrix
elements will here be made for off-mass shell vector mesons
with zero energy. This choice of kinematics is made
with application of resonance propagation in nuclear
matter in mind. The coupling of spin-isospin modes
that propagate in matter to the $P-$shell resonances
has recently been shown to be both significant and
intricate \cite{Peters,Friman,Friman1}. A key aim of the 
present study is to determine
the strength of this coupling. 

Only rough correspondences can be made between the couplings
in the the chiral quark model and in the hadronic model. 
Nevertheless, the coherence in the $\omega-$meson
coupling to the nucleon, where the factor 3 in the
$SU(3)$ relation $g_{\omega NN}\simeq 3 g_{\rho NN}$
arises in the sum over the three quarks in the
nucleon, seems to disappear as the coupling 
$g_{\omega NN^*}^{(1520)}$ in the quark model is
roughly equal to $g_{\rho NN}$ in the hadronic model,
which lacks coherence in the quark sum. In the quark
model we find that $g_{\omega NN^*}^{(1520)}
\sim 1/2 g_{\rho NN^*}^{(1520)}$. The same ratio of
$\omega$ to $\rho$ transition couplings is found in
hadronic resonance models as discussed below. A simple
explanation for this is still wanting.

This paper is divided into 4 sections. In section 2 we derive the
pion-resonance couplings and compare them to empirical data. The
$\rho$-meson and $\omega$-meson resonance transition couplings are
derived in section 3. A summarizing discussion is given in section 4.\\

\section{Pion coupling constants}

Before deriving the expressions for the vector meson transition
couplings to nucleon resonances it is instructive to
derive the corresponding pion transition coupling
constants within the quark model. Under the assumption
that the pion transitions are described by single quark
operators one may also derive these coupling strengths
directly from experiment. Overall the single quark 
operator approximation for the pion coupling to nucleon
resonances underestimates the pion decay widths of the
resonances, and therefore mainly has qualitative value
\cite{Kra,Glozc}. It does however determine
the phases of the coupling
constants, and, as shown below, if these are multiplied by 
about a factor
2, the decay widths are within a few ten percent of
the empirical values.

The coupling of pions to constituent $u$ and $d$ quarks may be described
by the pseudovector coupling
$${\cal L}_{\pi qq}=i{f_{\pi qq}\over m_\pi}\bar
\psi_q\gamma_5\gamma_\mu\partial_\mu\vec \phi_\pi\cdot \vec
\tau \psi_q.\eqno(2.1)$$
Here $\vec \phi_\pi$ is the isovector pion field operator,
$\psi_q$ is the constituent quark field and $m_\pi$ is the pion mass.
The pseudovector pion-quark coupling constant
may be determined from the corresponding pion-nucleon coupling constant
by comparison to the $\pi NN$ coupling:
$${\cal L}_{\pi NN}=i{f_{\pi NN}\over
m_\pi}\bar\psi_N\gamma_5\gamma_\mu\partial_\mu \vec \phi_\pi\cdot \vec
\tau \psi_N.\eqno(2.2)$$
Comparison of the matrix elements of the Lagrangian (2.1) and (2.2) for the
case of a proton with spin up, using the $SU(6)$ quark model wave
functions in the case of the former, yields
$$<p,{1\over 2} |{\cal L}_{\pi NN}|p,{1\over 2} >=-i f_{\pi NN}
{k_3\over m_\pi}$$
$$<p,{1\over 2}|\sum_{q=1}^{3}{\cal L}_{\pi qq}|p,{1\over 2}>
=-i{5\over 3}f_{\pi qq}{k_3\over m_\pi},\eqno(2.3)$$
where $k_3$ is the third component of the pion momentum. This gives
the relation
$$f_{\pi qq}={3\over 5}f_{\pi NN}.\eqno(2.4)$$
As $f_{\pi NN}\simeq 1$ it follows that $f_{\pi qq}\simeq 0.6$. This
result for the pion-quark coupling constant is close to that, which is
obtained from the Goldberger-Treiman relation for quarks:
$$f_{\pi qq}={g_A^q\over 2}{m_\pi\over f_\pi}.\eqno(2.5)$$
With $g_A^q\simeq 0.88$ and the value $f_\pi=93$ MeV for the pion decay
constant this relation gives $f_{\pi qq}\simeq 0.66$.
This coupling model does give a reasonably satisfactory
account of the pion decay widths of the $D-$meson resonances,
for which the single quark current model should be adequate
\cite{Goity}.\

The chiral quark model may be employed to express the transition
coupling constant of pions to nucleon resonance in terms of the pion
quark coupling constant $f_{\pi qq}$. Since this is given in terms of
the pion-nucleon coupling  constant $f_{\pi NN}$ by (2.3), it thus
becomes possible to express all pion-resonance transition coupling
constants in terms of the pion-nucleon coupling constant. 

These relations will depend on the orbital wave functions of the 3
constituent quarks that form the baryons. We shall here use the
covariant harmonic oscillator model for the mass operator of the 3
quarks constructed in ref. \cite{Coe1} to generate the 3 quark wave functions.
That model is formed by an, in
effect, linear confining interaction with a spin, flavor and orbital
angular momentum dependent hyperfine interaction. It describes the
baryon spectrum up to $\simeq $ 1700 MeV to an accuracy of a few
percent, which is quite adequate for the present application.

The 3-quark wave functions will be bilinear combinations of spin-flavor
and orbital wave functions. The latter are products of harmonic
oscillator functions of the two Jacobi coordinates of the 3-quark system:
$$\vec r={1\over \sqrt{2}}(\vec r_1-\vec r_2),\eqno(2.6a)$$
$$\vec \rho=\sqrt{{2\over 3}}(\vec r_3-{\vec r_1+\vec r_2\over 2}).
\eqno(2.6b)$$
The 3-quark wave functions in the model of ref. \cite{Coe1} are
eigenfunctions of the mass operator
$${\cal M}_0=\sqrt{3(\vec \kappa^2+\vec k^2+\omega^4(\vec \rho^2+\vec
r^2))},\eqno(2.7)$$
where $\omega$ is a parameter that determines the strength of the
confining interaction. Jacobi momenta $\vec \kappa$ and $\vec k$ are the
canonical momentum operators that are conjugate to the Jacobi
coordinates (2.5). The numerical value for the oscillator parameter
$\omega$ is 311 MeV, as determined from the baryon spectrum.
This value will be used here. In the impulse approximation
the empirical value of the rms radius of the proton (0.8 fm) obtains
with $\omega=245$ MeV .

The eigenfunctions of the mass operator (2.6) are products of harmonic
oscillator functions of $\vec \rho$ and 
$\vec r$: $\varphi_{n_1l_1m_1}(\vec
\rho\,)\varphi_{n_2l_2m_2}(\vec r\,)$. The appropriate symmetrized
combinations of these with spin and isospin wave functions for the $S-$,
$P-$, $D-$ and lowest excited $S-$state resonances are listed in Table 1.
In the table $|{1\over 2},s>_\pm$ and $|{1\over 2},t>_\pm$ denote spin and
isospin wave functions of mixed symmetry, which are symmetric
$(+)("(112)")$ or antisymmetric $(-)("(121)")$ under exchange of
the spins or isospins of the first
two quarks. The states $|{3\over 2},s>$ and $|{3\over 2},t>$ denote spin
and isospin states with total spin and isospin ${3\over 2}$, which
are symmetric under exchange of any set of two coordinates.

The explicit expressions for the states of mixed symmetry with
spin-$z$ projection $s_z$ are
$$|{1\over 2},s_z>_+ =\sum_{abcm}({1\over 2},{1\over 2},a,b
\vert 1,m)({1\over 2},1,c,m\vert {1\over 2},s_z)
\vert a,b,c>,\eqno(2.8a)$$
$$|{1\over 2},s_z>_- =\sum_{ab}({1\over 2},{1\over 2},a,b
\vert 0,0)\vert a,b,s_z>.\eqno(2.8b)$$
Here $\vert a,b,c>$ represent product states of three
spins with $z-$projections $a$,$b$ and $c$ respectively.
The corresponding isospin states with isospin-$z$ projection 
$t_z$ are readily
constructed by analogy \cite{Close}.
The rule for construction of symmetric combinations of
product states of spin, isospin and spatial wave functions
is based on the outer products of $S_3$.

Given the 3 quark wave functions for the nucleon resonances in Table 1,
it becomes possible to calculate the matrix elements,
$$<p,{1\over 2}|\sum_{q=1}^{3}{\cal L}_{\pi qq}|N^{*+},{1\over
2}>=-i{f_{\pi qq}\over m_\pi}<p,{1\over
2}|\sum_{q=1}^{3}\vec \sigma^q\cdot (\vec k
-\omega_\pi\vec v_q)\tau_3^qe^{-i\vec k\cdot \vec
r_q}|N^{*+},{1\over 2}>,\eqno(2.9)$$
of the pion-quark Lagrangian (2.1). Here $N^*$ represents a nucleon or a
$\Delta$ resonance, and $\vec k$ is the momentum and 
$\omega_\pi$ the energy of the pion.
The velocity operator for quark $q$ is denoted
$\vec v_q$. These matrix
elements, with the overall factor $-if_{\pi qq}/m_\pi$ divided out, are
listed in Table 2 for the resonances in Table 1.

In Table 2 the orbital matrix elements have been calculated with the
eigenfunctions of the mass operator (2.7). These are harmonic oscillator
functions, with the explicit forms
$$\varphi_{000}(\vec \rho\,)=({\omega^2\over \pi})^{3/4}e^{-\rho^2
\omega^2/2},$$
$$\varphi_{01m}(\vec \rho\,)=\sqrt{2}\omega \vec \rho_m \varphi_{000}
(\vec \rho\,),$$
$$\varphi_{200}(\vec \rho\,)=\sqrt{{2\over 3}}\omega^2
(\rho^2-{3\over 2\omega^2})\varphi_{000}
(\vec \rho\,),$$
$$\varphi_{02m}(\vec \rho\,)=\sqrt{{4\over
15}}\omega^2\rho^2\varphi_{000}(\vec \rho\,)\sqrt{4\pi}Y_{2m}(\hat
\rho).\eqno(2.10)$$

In the Table 2 values of the following overlap integrals
have been used:
$$(\varphi_{000}(\vec \rho\,),e^{-i\sqrt{{2\over 3}}\vec \rho
\cdot \vec k}\varphi_{000}(\vec \rho\,))=
e^{-k^2/6\omega^2},\eqno(2.11a)$$
$$(\varphi_{000}(\vec \rho\,),e^{-i\sqrt{{2\over 3}}\vec \rho\cdot \vec
k}\varphi_{200}(\vec \rho\,))=
-{\sqrt{6}\over 18}{k^2\over \omega^2}e^{-k^2/
6\omega^2},\eqno(2.11b)$$
$$(\varphi_{000}(\vec \rho\,),e^{-i\sqrt{{2\over 3}}\vec \rho\cdot \vec
k}\varphi_{01m}(\vec \rho\,))=-i{\sqrt{3}\over 3}{k_m\over \omega}
e^{-k^2/6\omega^2},\eqno(2.11c)$$
$$(\varphi_{000}(\vec \rho\,),e^{-i\sqrt{{2\over 3}}\vec \rho\cdot \vec
k}\varphi_{02m}(\vec \rho\,))=-{\sqrt{3}\over 9}{k^2\over
\omega^2}e^{-k^2/6\omega^2}\sqrt{{4\pi\over 5}}Y_{2m}
(\hat k).\eqno(2.11d)$$

The goal here is not, however, these quark model relations per $se$,
but expressions for the pion transitions couplings to the resonances,
when these are described as (generalized) Rarita-Schwinger field
operators \cite{Cheng}. These couplings may be described by the following
Lagrangians:
$${\cal L}_{\pi N\Delta}^{(1232)}={f_{\pi N \Delta}^{(1232)}\over m_\pi}
\bar \psi
\chi^\dagger \partial_\mu\vec \phi\cdot \vec \chi\psi_\mu+h.c.,
\eqno(2.12a)$$
$${\cal L}_{\pi NN^*}^{(1440)}=i{f_{\pi NN^*}^{(1440)}\over m_\pi}\bar
\psi \chi^\dagger\gamma_5\gamma_\mu\partial_\mu\vec \phi\cdot \vec
\tau\chi\psi_{N^*}+h.c.,\eqno(2.12b)$$
$${\cal L}_{\pi N\Delta^*}^{(1600)}={f_{\pi N\Delta^*}^{(1600)}\over
m_\pi}\bar \psi \chi^\dagger\partial_\mu \vec \phi\cdot \vec \chi
\psi_\mu+h.c.,\eqno(2.12c)$$
$${\cal L}_{\pi NN^*}^{(1535)}=i{f_{\pi NN^*}^{(1535)}\over m_\pi}\bar
\psi \chi^\dagger \gamma_\mu\partial_\mu\vec \phi\cdot \vec
\tau\chi\psi_{N^*}+h.c.,\eqno(2.12d)$$
$${\cal L}_{\pi NN^*}^{(1520)}={f_{\pi NN^*}^{(1520)}\over m_\pi}\bar
\psi\chi^\dagger \gamma_5\partial_\mu\vec \phi\cdot \vec \tau
\chi\psi_\mu+h.c.,\eqno(2.12e)$$
$${\cal L}_{\pi N\Delta^*}^{(1620)}=i{f_{\pi N\Delta^*}^{(1620)}
\over m_\pi}\bar
\psi \chi^\dagger\gamma_\mu\partial_\mu\vec\phi\cdot \vec \chi
\psi_{\Delta^*},\eqno(2.12f)$$
$${\cal L}_{\pi N\Delta^*}^{(1700)}={f_{\pi NN^*}^{(1700)}\over
m_\pi}\bar \psi\chi^\dagger\gamma_5\partial_\mu\vec \phi\cdot 
\vec \chi\psi_\mu +h.c.,\eqno(2.12g)$$
$${\cal L}_{\pi NN^*}^{(1650)}=i{f_{\pi NN^*}^{(1650)}\over m_\pi}\bar
\psi\chi^\dagger\gamma_\mu\partial_\mu
\vec \phi\cdot \vec \tau\chi\psi_{N^*}+h.c.,\eqno(2.12h)$$
$${\cal L}_{\pi NN^*}^{(1700)}={f_{\pi NN^*}^{(1700)}\over m_\pi}\bar \psi
\chi^\dagger \gamma_5\partial_\mu\vec \phi\cdot \vec \tau \chi
\psi_\mu+h.c.,\eqno(2.12i)$$
$${\cal L}_{\pi NN^*}^{(1675)}={f_{\pi NN^*}^{(1675)}\over m_\pi^2}\bar
\psi\chi^\dagger\partial_\mu\partial_\nu\vec \phi\cdot
\vec \tau\chi\psi_{\mu\nu}+h.c,\eqno(2.12j)$$
$${\cal L}_{\pi NN^*}^{(1720)}={f_{\pi NN^*}^{(1720)}\over
m_\pi}\bar\psi \chi^\dagger\partial_\mu\vec\phi\cdot \vec
\tau\chi\psi_\mu+h.c.,\eqno(2.12k)$$
$${\cal L}_{\pi NN^*}^{(1680)}=i{f_{\pi NN^*}^{(1680)}\over
m_\pi^2}\bar\psi\chi^\dagger\gamma_5\partial_\mu\partial_\nu\vec\phi\cdot
\vec \tau\chi \psi_{\mu\nu}+h.c.\eqno(2.12l)$$
Here $\psi,\psi_\mu$ and $\psi_{\mu\nu}$ represent spin $1/2,3/2$ and spin
$5/2$ Rarita-Schwinger spinor fields respectively, and $\chi$ and
$\vec \chi$ represents isospin $1/2$ spinor and $3/2$ vector-spinors
respectively. 

The Rarita-Schwinger spinor field operators are defined as
$$\psi_\mu^M=\Sigma(1,{1\over 2},m,s\vert {3\over 2},M)
\epsilon_\mu(m)u_s,
\eqno(2.13a)$$
$$\psi_{\mu\nu}^M=\Sigma({3\over 2},1,r,n\vert{5\over 2}M)(1,{1\over
2},m,s\vert{3\over 2},r)\epsilon_\mu(m)\epsilon_\nu(n)u_s.\eqno(2.13b)$$
The spin $5/2$ spinor $\psi_{\mu\nu}$ is symmetric in the two 4-vector
indices. Above $u_s$ represents a Dirac spinor with $s_z=s$.

The resonance couplings (2.12) have been written in the chiral symmetry
mandated form, which requires that the couplings vanish with pion
4-momentum. For baryons on their mass shell the couplings to the
negative parity resonances may be simplified by use of the Dirac and
corresponding Rarita-Schwinger equations.

The matrix elements of the transition couplings (2.12) between resonance
states with charge $+e$ and spin-$z$ projection $+{1\over 2}$ and the
proton with spin up are listed in Table 3.
Direct comparison with the quark model couplings is possible only for
those terms in (2.12) that have the corresponding dependence on pion
momentum. Comparison of the terms, which depend on pion energy, would
require the corresponding quark model couplings to be spelled out
explicitly. 

With this qualification comparison of the matrix elements in
Tables 2 and 3 yield the sought for
expressions for the resonance transition couplings to pions, $f_{\pi
NN^*}$, in terms of the pion-quark coupling constant, $f_{\pi qq}$,
and then by (2.4) in
terms of the $\pi N$ pseudovector coupling constant $f_{\pi NN}$. The
resulting expressions are listed in Table 4. 
In order to have real coupling constants in the
Rarita-Schwinger formalism the phase factors $(-i)^l$ that appear
in the quark model matrix elements have been dropped.
These calculated values for
the $\pi NN^*$ coupling constants may be compared to the corresponding
values determined from the empirically known widths for $N\pi$ decay of
these resonances.

The empirical decay widths for $N\pi$ decay of the positive parity
$\Delta$ resonances $\Delta(1232)$ and $\Delta(1600)$ are obtained as
$$\Gamma={1\over 3}{f_{\pi N\Delta}^2\over 4\pi}{E'+m_N\over
m_\Delta}{k^3\over m_\pi^2}\eqno(2.14)$$
Here $E'$ is the energy of the final nucleon.
Insertion of the empirical decay widths and the corresponding
kinematical factors then yields the value $f_{\pi
N\Delta}^{(1232)}=2.2\pm 0.04$ and $f_{\pi N\Delta}^{(1600)}=0.51 \pm
0.07$. These values exceed the quark model values (1.55 and 0.47) in
Table 4 by factors 1.5 and 1.1 respectively. These underestimates are 
typical of the quark
model for the pion decays in the single quark approximation.

For the $N(1440)$ the decay width for $N(1440)\rightarrow N\pi$ is
obtained as
$$\Gamma=3{(f_{\pi NN^*}^{(1440)})^2\over 4\pi}{E'-m_N\over m^*}{k\over
m_\pi^2}(m^*+m_N)^2\eqno(2.15)$$
Here $m^*$ is the resonance mass. From the empirical decay width
for $N\pi$ decay $227 \pm 0.65$ MeV of the $N(1440)$ one
obtains $f_{\pi NN^*}^{(1440)}=0.39 \pm 0.06$. In this case the quark
model value 0.26 again represent an underestimate of about a 
factor 1.5. This
underestimate is a consequence of the fact that the
$\pi NN(1440)$ coupling vanishes at $k=0$ in the
quark model.

The $N\pi$ decay width for the spin ${1\over 2}^-$ resonances are
obtained as
$$\Gamma={f_{\pi NN^*}^2\over 4\pi}{E'+m_N\over m^*}
{k\over m_\pi^2}(m^*-m_N)^2.\eqno(2.16)$$
Here the factor $\alpha$ is 3 for isospin $1/2$ resonances and 1 for
isospin $3/2$ resonances with spin ${1\over 2}^-$. For the $N(1535)$,
$N(1650)$ and the $\Delta(1620)$ this expression yields the values
$f_{\pi NN^*}^{(1535)}=0.36 \pm 0.05$, $f_{\pi NN^*}^{(1650)}=0.31 \pm
0.03$ and $f_{\pi N\Delta}^{(1620)}=0.34 \pm 0.06$.
The quark model results for these coupling constants in Table 4 are 
within 30\% of these values.

The $N\pi$ decay widths for the spin ${3\over 2}^-$ resonances 
are obtained as
$$\Gamma={1\over 3}{f_{\pi NN^*}^2\over 4\pi}{E'-m_N\over
m^*}{k^3\over m_\pi^2}.\eqno(2.17)$$
From this expressions and the empirical $N\pi$ decay widths we obtain
the values $f_{\pi NN^*}^{(1520)}=1.56 \pm 0.06$, $f_{\pi
NN^*}^{(1700)}=0.36$ and $f_{\pi N\Delta}^{(1700)}=1.31$. The quark
model values in Table 4 are fairly close to the first and the last
of these values,
but falls below that for the $N(1700)$. 

Finally the width for the decay $N(1675)\rightarrow N\pi$ is obtained as
$$\Gamma ={2\over 5}{(f_{\pi NN^*}^{(1675)})^2
\over 4\pi}{E'+m_N\over m^*}{k^5\over m_\pi^4}.
\eqno(2.18)$$
The empirical decay width fraction $\sim 67$ MeV yields 
the value $f_{\pi
NN^*}^{(1675)}=0.10$. This is close to the
corresponding quark model value (0.09) in Table 4. Note that the empirical
decay width does not determine the phase of the pion resonance
transition coupling constants.

The expressions for the pionic decay widths of the
$D-$shell resonances $N(1720)$ and $N(1680)$ are \cite{Cheng}
$$\Gamma(N(1720)\rightarrow N\pi)={1\over 3}
{(f_{\pi NN^*}^{(1720)})^2\over 4\pi} {E'+m_N\over m^*}{k^3
\over m_\pi^2},\eqno(2.19)$$
$$\Gamma(N(1680)\rightarrow N\pi)={2\over 5}
{(f_{\pi NN^*}^{(1680)})^2\over 4\pi} {E'-m_N\over m^*}{k^5
\over m_\pi^4},\eqno(2.20)$$
respectively.
The fractional widths for $N\pi$ decay of these two resonances
are $22\pm 10$ MeV and $84\pm 10$ MeV respectively \cite{PDG}.
Given these widths we obtain the following coupling constant
magnitudes: $\vert (f_{\pi NN^*}^{(1720)} \vert \sim 0.25\pm 0.06$
and $\vert (f_{\pi NN^*}^{(1680)} \vert \sim 0.42\pm 0.04$.
Comparison of the calculated coupling constants in Table 4
shows that the quark model, in the present approximation,
overestimates the former one of
these coupling constants by about a factor 5 and underestimates
the latter one by almost a factor 4. This problem with the
pion decays of the $D-$shell resonances has been noted 
before \cite{Plex}.

The overall situation that emerges is that with the one-quark 
transition operators,
the quark model mostly underpredicts the resonance transition
couplings by factors 1 - 1.5 with the present wave function
model. The conclusion is that two-quark operators 
have to be significant for the
description of pionic transitions of the baryon resonances
\cite{Buch}. If neglected these have to be compensated for 
by multiplication of the
$\pi N N^*$ couplings by factors of the order 2-3.

\section{Vector meson coupling constants}

A universal $SU(2)$ symmetric model for the vector meson couplings to
constituent quarks would be the following:
$${\cal L}_{Vqq}=ig_{\rho qq}\bar\psi\gamma_\mu\vec \tau\cdot \vec
\rho_\mu\psi+ig_{\omega qq}\bar\psi \gamma_\mu
\omega_\mu\psi.\eqno(3.1)$$
Here $\vec \rho_\mu$ and $\omega_\mu$ are the $\rho$-meson and
$\omega$-meson field operators respectively. 

The vector meson coupling constant $g_{Vqq}$ may be determined from
either the $\omega$- or $\rho$-nucleon coupling constants by writing the
vector meson-nucleon coupling in the conventional form
$${\cal L}_{VNN}=ig_{\omega NN}\bar
\psi_N[\gamma_\mu+i{\kappa_\omega\over 2m}\sigma_{\mu\nu}\partial_\nu]
\omega_\mu\psi_N$$
$$+ig_{\rho NN}\bar \psi_N[\gamma_\mu+i{\kappa_\rho\over
2m}\sigma_{\mu\nu}\partial_\nu]\vec \rho_\mu\cdot \vec \tau\psi_N.
\eqno(3.2)$$
Comparison of the matrix elements of the charge components of these
Lagrangians for e.g. protons with spin up to the same matrix elements of
the quark coupling operator (3.2) yields the relations
$$g_{\omega NN}=3g_{Vqq},\eqno(3.3a)$$
$$g_{\rho NN}=g_{Vqq}.\eqno(3.3b)$$

The tensor couplings $\kappa_\omega$ and $\kappa_\rho$ in (3.3) may be
determined by comparing the matrix elements of the transverse part of
the current couplings in (3.1) and (3.3). This yields the relations
$${g_{\omega qq}\over m_q}={1\over m_N}g_{\omega NN}(1+
\kappa_\omega),\eqno(3.4a)$$
$${5\over 3}{g_{\rho qq}\over m_q}={1\over m_N}g_{\rho NN}(1+
\kappa_\rho).\eqno(3.4b)$$
Boson exchange models for the nucleon-nucleon interaction indicate that
$\kappa_\omega$ is small. From (3.4a) it then follows that
$m_q=m_N/3=313$ MeV, in agreement with conventional quark model
phenomenology. Equations (3.3b) and (3.3b) then imply that
$\kappa_\rho=4$. This is close to the value $\kappa_\rho=4.22$ in a recent
realistic boson exchange model for the nucleon-nucleon
interaction \cite{Nijm}, but somewhat smaller than the value 6.6
indicated in earlier interaction models \cite{Bonn}. The values for the
$\omega NN$ coupling constants differ between different potential
models. In the recent Nijmegen model \cite{Nijm} it is $g_{\omega
NN}=10.35$, while in the Bonn model it is as big as $g_{\omega
NN}=15.85$. The values for the $\rho NN$ vector coupling constant are
more stable, ranging from $g_{\rho NN}=2.97$  \cite{Nijm} to 3.19
\cite{Bonn}. These uncertainties are commensurate with the expected
uncertainties in the quark model. 

These numbers are consistent with assuming equality between the
$\rho$ and $\omega$ couplings to constituent quarks, and with
taking
$$g_{\rho qq}=g_{\omega qq}\simeq 3.\eqno(3.5)$$
The anomalous tensor couplings of the constituent quarks 
are so small that they may be taken to be 0.
We shall use these values here.

In Table 5 the explicit matrix elements of the charge and
transverse current components of the $\omega-$quark
coupling (3.1) are given for all the nucleon and
$\Delta$ resonances in Table 1 are listed. Here only
the non-vanishing terms of lowest order in the
$v/c$ expansion have been included. The corresponding
matrix elements of the $\rho-$quark coupling are listed
in Table 6.

As the aim here is to calculate the vector meson transition
couplings to nucleon resonances by means of the quark
model, the effective coupling Lagrangians will have to
be expressed in a form, which has the same momentum
dependence as the corresponding transition matrix
elements in the quark model. The standard form of the
generalized Rarita-Schwinger vector current couplings
in ref. \cite{Cheng} does not meet this criterion. The
form of the transition couplings below have been chosen
to have have the same momentum dependence as the quark
model couplings. As a consequence the overall meson
momentum factors drop out in the expressions for the
transition couplings in terms of the corresponding
vector meson coupling constants to nucleons.

The $\omega$-meson transition couplings to the nucleon resonances may 
be described by the following effective Lagrangians:
$${\cal L}_{\omega NN^*}^{(1440)}=-i{g_{\omega NN^*}
^{(1440)}\over m_\omega^2}\bar \psi_N
[\gamma_\mu-{m^*-m_N\over
m_\omega^2}\partial_\mu]\partial^2\omega_\mu\psi_{N^*}+h.c.,\eqno(3.6a)$$
$${\cal L}_{\omega NN^*}^{(1535)}=-i{g_{\omega NN^*}^{(1535)}
\over m_\omega^2}\bar \psi_N
\gamma_5[\gamma_\mu\partial^2-(m^*+m_N)\partial_\mu]\omega_\mu
\psi_{N^*}+h.c.,\eqno(3.6b)$$
$${\cal L}_{\omega NN^*}^{(1520)}=i{g_{\omega NN^*}^{(1520)}
\over m_\omega^2}\bar\psi_N\sigma_{\mu\nu}\partial_\nu\partial_\kappa
\omega_\mu\psi_\kappa+h.c.,\eqno(3.6c)$$
$${\cal L}_{\omega NN^*}^{(1650)}=-i{g_{\omega NN^*}^{(1650)}
\over m_\omega^2}\bar
\psi_N\gamma_5[\gamma_\mu\partial^2-(m^*+m_N)\partial_\mu]
\omega_\mu\psi_{N^*}+h.c.,\eqno(3.6d)$$
$${\cal L}_{\omega NN^*}^{(1700)}=i{g_{\omega NN^*}^{(1700)}
\over m_\omega^2}\bar\psi_N\sigma_{\mu\nu}
\partial_\nu\partial_\kappa\omega_\mu
\psi_\kappa+h.c.,\eqno(3.6e)$$
$${\cal L}_{\omega NN^*}^{(1675)}=i{g_{\omega
NN^*}^{(1675)}\over m_\omega^2}
\epsilon_{\mu\alpha\beta \delta}\bar\psi_N\partial_\alpha\partial_\nu
\omega_\beta\gamma_\delta\psi_{\mu\nu}+h.c.,\eqno(3.6f)$$
$${\cal L}_{\omega NN^*}^{(1720)}={g_{\omega NN^*}
^{(1720)}\over m_\omega^2}\bar \psi_N\gamma_5
[\delta_{\mu\nu}-{1\over m^*+m_N}\gamma_\mu\partial_\nu]
\partial^2\omega_\mu\psi_\nu+h.c.,\eqno(3.6g)$$
$${\cal L}_{\omega NN^*}^{(1680)}=i{g_{\omega NN^*}
^{(1680)}\over m_\omega^2}\bar \psi_N
[\gamma_\mu-{m^*-m_N\over
m_\omega^2}\partial_\mu]\partial_\alpha
\partial_\beta\omega_\mu\psi_{\alpha\beta}+h.c..\eqno(3.6h)$$

The matrix element of the charge and transverse current coupling terms
of these Lagrangians between resonances with charge state $+e$ and the
proton with spin up are listed in Table 7. As these matrix elements
relate to virtual vector meson production, we have here 
considered the vector mesons as having zero energy.
For the comparison with the quark
model operators, we have in
addition dropped terms of order $(m^*-m_N)/(m^*+m_N)$. For the heavier
resonances this introduces a theoretical uncertainty range of almost
30\%, which however is inherent in the mismatch between 
the non-relativistic
quark model expressions and 
the covariant Rarita-Schwinger formalism. 
Numerical estimates for the $\omega$ meson transition couplings
to the nucleon resonances are given in Table 8. In the
calculation of the numerical values, the quark
wave function factors exp$\{-\vec k^2/6\omega^2\}$ were set
to unity.

The $\rho$-meson transition coupling to the nucleon and
$\Delta$-resonances are described by the following coupling Lagrangians
in the generalized Rarita-Schwinger formalism:
$${\cal L}_{\rho
N\Delta}^{(1232)}=g_{\rho N\Delta}^{(1232)}\bar\psi\chi^\dagger
\gamma_5[\delta_{\mu\nu}-{1\over
m_\Delta+m_N}\gamma_\mu\partial_\nu]\vec \rho_\mu\cdot \vec \chi
\psi_\nu+h.c.,\eqno(3.7a)$$
$${\cal L}_{\rho NN^*}^{(1440)}=-i{g_{\rho
NN^*}^{(1440)}\over m_\rho^2}\bar\psi\chi^\dagger[\gamma_\mu-{m^*-m_N\over
m_\rho^2}\partial_\mu+i{\kappa_{\rho NN}^{(1440)}\over
m^*-m_N}\sigma_{\mu\nu}\partial_\nu]\partial^2
\vec \tau\cdot \vec \rho_\mu
\chi\psi_{N^*}$$
$$+h.c.,\eqno(3.7b)$$
$${\cal L}_{\rho N\Delta^*}^{(1600)}=-g_{\rho N\Delta^*}^{(1600)}\bar
\psi\chi^\dagger\gamma_5[\delta_{\mu\nu}-{1\over
m_{\Delta^*}+m}\gamma_\mu\partial_\nu]\partial^2\vec \rho_\mu\cdot \vec
\chi\psi_\nu+h.c.,\eqno(3.7c)$$
$${\cal L}_{\rho NN^*}^{(1535)}=-i{g_{\rho NN^*}^{(1535)}\over
m_\rho^2}\bar
\psi\chi^\dagger\gamma_5[\gamma_\mu\partial^2-(m^*+m_N)
\partial_\mu]\vec \tau\cdot \vec \rho_\mu\chi
\psi_{N^*}+h.c.,\eqno(3.7d)$$
$${\cal L}_{\rho NN^*}^{(1520)}= i{g_{\rho NN^*}^{(1520)}
\over m_\rho^2}
\bar\psi\chi^\dagger\sigma_{\mu\nu}\partial_\nu\partial_\kappa
 \vec \tau\cdot \vec \rho_\mu
\chi\psi_\kappa+h.c.,\eqno(3.7e)$$
$${\cal L}_{\rho N\Delta^*}^{(1620)}=-i{g_{\rho
N\Delta^*}^{(1620)}\over m_\rho^2}\bar\psi\chi^\dagger
\gamma_5[\gamma_\mu\partial^2-
(m_{\Delta^*}+m_N)\partial_\mu]\vec
\rho_\mu\cdot \vec \chi\psi_{\Delta^*}+h.c.,\eqno(3.7f)$$
$${\cal L}_{\rho N\Delta^*}^{(1700)}=i{g_{\rho
N\Delta^*}^{(1700)}
\over m_\rho^2}
\bar\psi\chi^\dagger\sigma_{\mu\nu}\partial_\nu\partial_\kappa
\vec \rho_\mu\cdot
\vec \chi\psi_\kappa+h.c.,\eqno(3.7g)$$
$${\cal L}_{\rho NN^*}^{(1650)}=-i{g_{\rho NN^*}^{(1650)}\over
m_\rho^2}\bar\psi
\chi^\dagger \gamma_5[\gamma_\mu\partial^2-(m^*+m_N)\partial_\mu]\vec
\tau\cdot \vec \rho_\mu\chi \psi_N+h.c.,\eqno(3.7h)$$
$${\cal L}_{\rho NN^*}^{(1700)}=i{g_{\rho NN^*}^{(1700)}
\over m_\rho^2}\bar\psi
\chi^\dagger \sigma_{\mu\nu}\partial_\nu\partial_\kappa
\vec
\tau\cdot \vec \rho_\mu\chi \psi_\kappa+h.c.,\eqno(3.7i)$$
$${\cal L}_{\rho NN^*}^{(1675)}=i{g_{\rho NN^*}^{(1675)}
\over m_\rho^2}\epsilon_{\mu\alpha\beta\delta}\bar\psi_N\chi^\dagger
\partial_\alpha\partial_\nu\vec \tau\cdot \vec \rho_\beta\gamma_\delta
\chi \psi_{\mu\nu}+h.c.,\eqno(3.7j)$$
$${\cal L}_{\rho NN^*}^{(1720)}={g_{\rho NN^*}
^{(1720)}\over m_\rho^2}\bar \psi_N\chi^\dagger\gamma_5
[\delta_{\mu\nu}-{1\over m^*+m_N}\gamma_\mu\partial_\nu]
\partial^2\vec\tau\cdot\vec \rho_\mu\chi\psi_\nu+h.c.,\eqno(3.7k)$$
$${\cal L}_{\rho NN^*}^{(1680)}=i{g_{\rho NN^*}
^{(1680)}\over m_\rho^2}\bar \psi_N\chi^\dagger
[\gamma_\mu-{m^*-m_N\over
m_\rho^2}\partial_\mu]\partial_\alpha
\partial_\beta\vec \tau\cdot\vec\rho_\mu\chi
\psi_{\alpha\beta}+h.c..\eqno(3.7l)$$

The matrix elements of these transition couplings are
listed in Table 9 for resonances with charge state $e$ and
protons with spin $1/2$. The explicit expressions for the 
$\rho-$meson transition couplings in terms of the 
corresponding $\rho NN$ coupling constant are listed
in Table 10, along with numerical estimates. These
expressions are obtained by comparison of the
quark model transition matrix elements in Table 6 with
the corresponding matrix elements of the transition
couplings in Table 9.

Of the resonances considered only the $N(1720)$ lies
above threshold for $N\rho$ decay, for which the
decay branch is in fact large ($\sim 80\%$). This makes
it possible to determine magnitude of the transition coupling
constant $g_{\rho NN^*}^{(1720)}$ directly from its decay
width. For a real $\rho$ meson the coupling Lagrangian
(3.7k) simplifies as the differential operator
$\partial^2$ may be replaced by $m_\rho^2$. To lowest
order in $v/c$ the Lagrangian (3.7c) then reduces to that
of ref. \cite{Pirner}, with the identification
$$f_{N^* N\rho}= -2{\mu\over m_\rho}g_{\rho NN^*}^{(1720)},
\eqno(3.8)$$
where $f_{N^* N\rho}$ is the transition coupling constant
defined in ref. \cite{Pirner}. This was determined from
the partial decay width for $N(1720)\rightarrow N\rho$ to be
$f_{N^* N\rho}\simeq 7.2$ in ref.\cite{Pirner}. This is
somewhat smaller than the value 13.8 that is obtained
from eqn.(3.8) with the value for $g_{\rho NN^*}^{(1720)}$
obtained with the quark model in Table 9. The fact that
the quark model leads to an overprediction for the
$\rho NN(1720)$ transition strength in clearly related
to the corresponding overprediction of the $\pi NN(1720)$
coupling constant noted above. 

\section{Discussion}

The fact that the pion resonance transition couplings are 
underestimated by factors 1.5--2 by the
single quark operator approximation suggests that the
quark model, in the same approximation, may also
lead to similar underestimates for the vector meson transition
couplings to nucleon resonances. For the subthreshold
resonance transition couplings considered here,
and which are required in a dynamical treatment of
nuclear matter, there is
however no alternative to calculation based on a dynamical 
model. 

The expressions for the vector meson transition coupling
constants were derived here by comparing the matrix
elements of the transverse components of the vector
meson transition currents to the corresponding quark
model matrix elements. The generalized Rarita-Schwinger coupling
Lagrangians in eqs. (3.6) and (3.7) are, however, invariant
and may be applied for virtual vector mesons of arbitrary
momentum and energy in nuclear matter. As an example,
consider the coupling (3.7e) of the $\rho-$meson to
the $N(1520)$ $3/2^-$ resonance, which admits an
interpretation as a $\rho-$nucleon resonance.
For non-zero $\rho-$meson energy this coupling, to lowest
order in the inverse baryon masses takes the form
$${\cal L}_{\rho N N^*}\simeq{g_{\rho NN^*}^{(1520)}\over 
m_\rho^2}\psi^\dagger\chi^\dagger\{{\vec k^2\over 2
\mu}
\rho_0^a+i(1-{\omega_\rho\over 2
\mu})\vec\sigma\cdot\vec k
\times\vec \rho ^a\}\vec \tau^a
\vec k\cdot \vec\psi\chi+h.c. \eqno(4.1)$$
Here $\mu$ is defined as the baryon mass
combination $\mu=2m^*m_N/(m^*+m_N)$. 

This form of the $\rho NN(1520)$ coupling, which takes
into account the $L=1$ aspect of the 3 quark description
of the $N(1520)$ resonance differs from the form commonly
employed for the same coupling \cite{Peters,Friman}:
$${\cal L}_{\rho N N^*}^{(1520)}={f_{\rho NN^*}^{(1520)}\over
m_\rho}\psi^\dagger\chi^\dagger(\omega\vec\rho^a-\rho_0^a 
\vec k)\tau^a\cdot\vec\psi\chi+h.c. \eqno(4.2)$$
If in (4.1) one sets $\vec k^2=k^2$, as appropriate for
zero energy vector mesons, and then imposes the on-shell
condition $k^2=-m_\rho^2$, along with the relation
$\vec\psi=i\vec\sigma\times\vec\psi$, 
a formal equivalence between the expressions (4.1)
and (4.2) obtains, provided that
$$f_{\rho NN^*}^{(1520)}={m_\rho\over 2\mu}g_{\rho NN^*}
^{(1520)}.\eqno(4.3)$$
Comparing numbers, with the value 4.5 for $g_{\rho NN^*}
^{(1520)}$ given in Table 10, we obtain
$f_{\rho NN^*}{(1520)}=1.5$, which is about half of the
value 3.2 obtained in ref.\cite{Friman}. Given the fact that
the quark model underestimates the pion resonance
transition couplings by factors 1.5--2 in the single
quark operator approximation, this small quark model
value is not unexpected.

	This comparison may also be extended to the case
of the $\omega N N(1520)$ transition coupling.
By comparing the isospin independent versions of the
transition Lagrangians (4.1) and (4.2), we obtain
$$f_{\omega NN^*}^{(1520)}={m_\omega\over 2\mu}
g_{\omega NN^*}
^{(1520)}.\eqno(4.4)$$
With the value $g_{\omega NN^*}^{(1520)}=7.7$
obtained by means of the quark model in Table 7, one
obtains $f_{\omega NN^*}^{(1520)}=2.6$. This value
is somewhat less than one half of that obtained
in ref.\cite{Friman}.
The smaller value may be a consequence of the fact that
the chiral quark model result should apply to
higher densities as discussed above. It roughly
corresponds to about $1/3$ of $g_{\omega NN}$, which
is the $\omega-$nucleon coupling at zero density. The
factor $1/3$ in the $SU(3)$ relation
$$g_{\omega NN}= 3 g_{\rho NN},\eqno(4.5)$$
arises from coherence in the sum of the couplings of the
three nonstrange quarks in the $\omega$. We would
expect this coherence to disappear at higher density or
temperature scales. We see no obvious simple reason for
why for this resonance $g_{\rho NN^*}$ is only about
$1/2$ of $g_{\omega NN^*}$, however.

This work should only be viewed as a first attempt at
calculating the transition couplings for virtual
vector mesons, and therefore the numerical values
obtained should be viewed as suggestive, rather
than as definite quantitative predictions. In an
earlier study \cite{GEB} we found that two-pion
exchange between constituent quarks furnished a
significant contribution to the hyperfine interaction
between constituent quarks, which, when combined with
the one-pion exchange interaction, provides a dynamical
basis for the effective spin-flavor structure that
is required for a satisfactory description of 
the empirical spectra, when combined with a linear
confining interaction.. 
A study of how such higher order corrections in the 
chiral quark model affects the problem at hand is now being 
undertaken. The goal is a better understanding of the
change from meson to quark variables, given the
general view that there is a region where the two
descriptions overlap \cite{Manoh}.

The present method for calculating
the resonance transition couplings to the nucleon
and $\Delta$ resonances in the $P-$ and $SD-$ shells
may be directly generalized to
the higher lying resonances. The 
explicit quark model wave functions for all the
resonances in the $SD-$shell may be constructed by
reference to the symmetry classification for the
higher resonances in ref.\cite{Gloza}. The construction
of the corresponding transition couplings in the
generalized Rarita-Schwinger formalism may be
carried out with the methods outlined in ref.\cite{Cheng}
once care is taken to match the momentum dependence
of the quark model matrix elements. 

\vspace{1.5cm}

\centerline{\bf Acknowledgement}

\vspace{0.5cm}

D. O. R. thanks the Physics Department of Brookhaven National
Laboratory for its hospitality during the completion of
this work.
Research supported in part by the U.S. Department of Energy under grant
DE-FG02-88ER40388 and the Academy of Finland by 
grants No. 43982 and 44903.

\begin{tabular}{|l|l|}\hline
\multicolumn{2}{|l|}{Table 1. Explicit wave functions for the nucleon
and $\Delta$ resonances in the $S$, $P$}\\
\multicolumn{2}{|l|}{and $D$ shells including the lowest
excited $S$-shell states.}\\ \hline
 & \\
$p,n,{1\over 2}^+$ & ${1\over \sqrt{2}}\varphi_{000}(\vec
\rho\,)\varphi_{000}(\vec r\,)
\{|{1\over 2},t_3>_+|{1\over 2},s_3>_++|{1\over
2},t_3>_-|{1\over 2},s_3>_-\}$\\
 & \\
$\Delta(1232),{3\over 2}^+$ & $\varphi_{000}(\vec
\rho\,)\varphi_{000}(\vec r\,)|{3\over 2},t_3>|{3\over 2},s_3>$\\
 & \\
$N(1440),{1\over 2}^+$ & ${1\over 2}\{\varphi_{200}(\vec
\rho\,)\varphi_{000}(\vec r\,)+\varphi_{000}(\vec \rho\,)
\varphi_{200}(\vec r\,)\}$\\
 & \\
  & $\{|{1\over 2},t_3>_+|{1\over 2},s_3>_+ +|{1\over 2},t_3>_-|
{1\over 2},s_3>_-\}$\\
 & \\
$\Delta(1600),{3\over 2}^+$ & ${1\over \sqrt{2}}
\{\varphi_{200}(\vec \rho\,)
\varphi_{000}(\vec r\,)+\varphi_{000}(\vec \rho\,)\varphi_{200}(\vec r\,)\}
|{3\over 2},t_3>|{3\over 2},s_3>$\\
 & \\
$N(1535), {1\over 2}^-$ & ${1\over 2}\sum_{ms}(1,{1\over
2},m,s|J,s_3)\{\varphi_{01m}(\vec \rho\,)\varphi_{000}(\vec r\,)$\\
 & \\
$N(1520), {3\over 2}^-$ & $[|{1\over 2},t_3>_+|{1\over 2},
s>_+-|{1\over 2},t_3>_-|{1\over 2},s>_-]$\\
 & \\
 & $-\varphi_{000}(\vec \rho\,)\varphi_{01m}(\vec r\,)[|{1\over 2},
t_3>_+|{1\over 2},s>_-+|{1\over 2},t_3>_-|{1\over 2}s>_+]\}$\\
 & \\
$\Delta (1620),{1\over 2}^-$ & ${1\over \sqrt{2}}\sum_{ms}(1,{1\over
2},m,s|J,s_3)\{\varphi_{01m}(\vec \rho\,)\varphi_{000}(\vec r\,)|{3\over
2},t_3>|{1\over 2},s>_+$\\
 & \\
$\Delta (1700),{3\over 2}^-$ & $+\varphi_{000}(\vec
\rho\,)\varphi_{01m}(\vec r\,)|{3\over 2},t_3>|{1\over 2}, s>_-\}$\\
 & \\
$N(1650),{1\over 2}^-$ & ${1\over \sqrt{2}}\sum_{ms}(1,{3\over
2},m,s|J,s_3)\{\varphi_{01m}(\vec \rho\,)\varphi_{000}
(\vec r\,)|{1\over 2},t_3>_+$\\
 & \\
$N(1700),{3\over 2}^-$ & $+\varphi_{000}(\vec \rho\,)\varphi_{01m}(\vec
r\,)|{1\over 2},t_3>_-\}|{3\over 2},s>$\\
 & \\
$N(1675),{5\over 2}^-$ & \\
 & \\
$N(1720),{3\over 2}^+$ & ${1\over 2}\sum_{ms}(2,{1\over
2},m,s|J,s_3)\{\varphi_{02m}(\vec \rho\,)\varphi_{000}(\vec r\,)
+\varphi_{000}(\vec \rho\,)\varphi_{02m}(\vec r\,)\}$\\
 & \\
$N(1680),{5\over 2}^+$ & $\{|{1\over 2},t_3>_+|{1\over 2},s_3>_+
+|{1\over 2},t_3>_-|{1\over
2},s_3>_-\}$ \\ 
& \\
\hline
\end{tabular} 

\vspace{1cm}

\begin{tabular}{|l|l|}\hline
\multicolumn{2}{|l|}{Table 2. Transition matrix elements of the quark
operator $O=\sum_{q=1}^{3}\vec \sigma^q\cdot \vec k\tau_3^qe^{-i\vec
k\cdot \vec r_q}$}\\
\multicolumn{2}{|c|}{between
the nucleon resonances and the nucleon for charge states $+1$ with
$s_z=+{1\over 2}.$}\\ \hline
 & \\
$<p,{1\over 2}|O|p,{1\over 2}>$ & ${5\over 3}k_3$\\
 & \\
$<p,{1\over 2}|O|\Delta(1232)^+,{1\over 2}>$ & ${4\sqrt{2}\over
3}k_3e^{-k^2/6\omega^2}$\\ \hline
 & \\
$<p,{1\over 2}|O|N(1440)^+,{1\over 2}>$ & ${5\sqrt{3}\over
54}k_3({k^2\over \omega^2}
+{3\over 2}{\omega_\pi\over m_q})
e^{-k^2/6\omega^2}$\\
 & \\
$<p,{1\over 2}|O|\Delta(1600)^+,{1\over 2}>$ & ${3\sqrt{6}\over
27}k_3({k^2\over \omega^2}
+{3\over 2}{\omega_\pi\over m_q}
)e^{-k^2/6\omega^2}$\\ \hline
 & \\
$<p,{1\over 2}|O|N(1535)^+,{1\over 2}>$ & $-i{2\sqrt{2}\over 9}
\omega({k^2\over \omega^2}+{9\omega_\pi\over 2 m_q})
e^{-k^2/6\omega^2}$\\
& \\
$<p,{1\over 2}|O|N(1520)^+,{1\over 2}>$ & $i{2\over 9}{3k_3^2-\vec
k^2\over \omega}e^{-k^2/6\omega^2}$\\ \hline
& \\
$<p,{1\over 2}|O|\Delta(1620)^+,{1\over 2}>$ & $i{\sqrt{2}\over
9}\omega({k^2\over \omega^2}+{9\omega_\pi\over 2 m_q})
e^{-k^2/6\omega^2}$\\
& \\
$<p,{1\over 2}|O|\Delta(1700)^+,{1\over 2}>$ & $-{i\over 3}{3k_3^2-\vec
k^2\over \omega}e^{-k^2/6\omega^2}$\\ \hline
& \\
$<p,{1\over 2}|O|N(1650)^+,{1\over 2}>$ & $-i{\sqrt{2}\over 9}
\omega({k^2\over \omega^2}+{9\omega_\pi\over 2 m_q})
e^{-k^2/6\omega^2}$\\
& \\
$<p,{1\over 2}|O|N(1700)^+,{1\over 2}>$ & $-i{\sqrt{10}\over
90}{3k_3^2-\vec k^2\over \omega}e^{-k^2/6\omega^2}$\\
 & \\
$<p,{1\over 2}|O|N(1675)^+,{1\over 2}>$ & $i{\sqrt{10}\over
15}{3k_3^2-\vec k^2\over \omega}e^{-k^2/6\omega^2}$\\ \hline
 & \\
$<p,{1\over 2}|O|N(1720)^+,{1\over 2}>$ & $-{\sqrt{15}\over
27}k_3({k^2\over \omega^2}
+{15\omega_\pi \over 2 m_q})
e^{-k^2/6\omega^2}$ \\
 & \\
$<p,{1\over 2}|O|N(1680)^+,{1\over 2}>$ & ${\sqrt{45}\over
108}{k_3(5k_3^2-3k^2)\over \omega^2}e^{-k^2/6\omega^2}$\\ \hline
\end{tabular}

\begin{tabular}{|ll|} \hline
\multicolumn{2}{|l|}{Table 3. Transition matrix elements $<p,{1\over
2}|{\cal L}_{\pi NN^*}|N^{*+},{1\over 2}>$ of the pion transition}\\
\multicolumn{2}{|l|}{couplings in eqs. (2.12). Here $m^*$ denotes the
mass of the corresponding resonance}\\
\multicolumn{2}{|l|}
{The expressions after the vertical bars correspond to zero
energy pions.}\\
 \hline
 & \\
$<p,{1\over 2}|{\cal L}_{\pi N\Delta}^{(1232)}|\Delta(1232)^+,{1\over
2}>$ & $-{2\over 3}i{f_{\pi N\Delta}\over m_\pi}k_3$\\ \hline
 & \\
$<p,{1\over 2}|{\cal L}_{\pi NN^*}^{(1440)}|N(1440)^+,{1\over 2}>$ &
$-i{f_{\pi NN^*}^{(1440)}\over m_\pi}k_3$\\
 & \\
$<p,{1\over 2}|{\cal L}_{\pi N\Delta^*}^{(1600)}|\Delta(1600)^+,{1\over
2}>$ & $-i{2\over 3}{f_{\pi N\Delta^*}^{(1600)}\over m_\pi}
k_3$\\ \hline
 & \\
$<p,{1\over 2}|{\cal L}_{\pi NN^*}^{(1535)}|N(1535)^+,{1\over 2}>$ &
$if_{\pi NN^*}^{(1535)}{m^*-m\over  m_\pi}\,\,\,\,\vert\,\,\,
\{{i\over 4}{f_{\pi NN^*}^{(1535)}\over m_\pi}{m^*+m\over m^*m}\vec k^2\}$
  \\ &\\
$<p,{1\over 2}|{\cal L}_{\pi NN^*}^{(1520)}|N(1520)^+,{1\over 2}>$ &
${i\over 4\sqrt{6}}{f_{\pi NN^*}^{(1520)}\over m_\pi}{m^*-m\over
m^*m}(3k_3^2-\vec k^2)$\\ \hline
 & \\
$<p,{1\over 2}|{\cal L}_{\pi N\Delta^*}^{(1620)}|\Delta(1620)^+,{1\over
2}>$ & $i\sqrt{{2\over 3}}f_{\pi N\Delta^*}^{(1620)}
{m^*+m\over  m_\pi}\,\,\,\,\vert\,\,\,
\{{i\over 4}\sqrt{{2\over 3}}{f_{\pi N\Delta^*}^{(1620)}\over m_\pi}
{m^*-m\over m^*m}\vec k^2\}$\\
 & \\
$<p,{1\over 2}|{\cal L}_{\pi N\Delta^*}^{(1700)}|\Delta(1700)^+,{1\over
2}>$ & ${i\over 12}{f_{\pi N\Delta^*}^{(1620)}\over m_\pi}
{m^*-m\over m^*m}(3k_3^2-\vec k^2)$ \\\hline
 & \\
$<p,{1\over 2}|{\cal L}_{\pi NN}^{(1650)}|N(1650)^+,{1\over 2}>$ &
$if_{\pi N N^*}^{(1650)}{m^*-m\over  m_\pi}\,\,\,\,\vert\,\,\,
\{{i\over 4}{f_{\pi NN^*}^{(1650)}\over m_\pi}{m^*+m\over m^*m}\vec k^2\}$\\
&\\
$<p,{1\over 2}|{\cal L}_{\pi NN^*}^{(1700)}|N(1700)^+,{1\over 2}>$ &
${i\over 4\sqrt{6}}{f_{\pi NN^*}^{(1700)}\over m_\pi}
{m^*-m\over m^*m}(3k_3^2-\vec k^2)$\\
 & \\
$<p,{1\over 2}|{\cal L}_{\pi NN^*}^{(1675)}|N(1675)^+,{1\over 2}>$ &
$-{i\over \sqrt{10}}{f_{\pi NN^*}^{(1675)}\over m_\pi}{(3k_3^2-\vec k^2)
\over m_\pi}$\\ \hline
 & \\
$<p,{1\over 2}|{\cal L}_{\pi NN^*}^{(1720)}|N(1720)^+,{1\over 2}>$ &
$-i\sqrt{{2\over 3}}{f_{\pi NN^*}^{(1720)}\over m_\pi}{k_3\over
m_\pi}$\\
 & \\
$<p,{1\over 2}|{\cal L}_{\pi NN^*}^{(1680)}|N(1680)^+,{1\over 2}>$ &
$i{1\over 4}\sqrt{{2\over 5}}{f_{\pi NN^*}^{(1680)}\over
m_\pi^2}{m^*+m\over m^*m}k_3(5k_3^2-3k^2)$\\ \hline
\end{tabular}

\begin{tabular}{|l|} \hline
Table 4. The resonance transition coupling constants in terms of the\\
pion-nucleon pseudoscalar coupling constant $f_{\pi NN}$.\\ \hline
 \\
$f_{\pi N\Delta}^{(1232)}={6\sqrt{2}\over 5}
e^{-k^2/6\omega^2}f_{\pi NN}\simeq 1.55 f_{\pi NN}$\\ \hline
 \\
$f_{\pi NN^*}^{(1440)}={\sqrt{3}\over 18}({k^2\over
\omega^2}
+{3\over 2}{\omega_\pi\over m_q})
e^{-k^2/6\omega^2}f_{\pi NN}\simeq 0.26 f_{\pi NN}$\\
 \\
 $f_{\pi N\Delta}^{(1600)}={\sqrt{6}\over 10}
({k^2\over \omega^2}
+{3\over 2}{\omega_\pi\over m_q})
e^{-k^2/6\omega^2}f_{\pi NN}\simeq 0.47 f_{\pi
NN}$\\ \hline
 \\
$f_{\pi NN^*}^{(1535)}={2\sqrt{2}\over 15}{\omega\over
(m^*-m_N)}
({k^2\over \omega^2}+{9\omega\over 2 m_q})
e^{-k^2/6\omega^2}f_{\pi NN}\simeq 0.49 f_{\pi NN}$\\
 \\
$f_{\pi NN^*}^{(1520)}=-{8\sqrt{6}\over 15}{m^* m_N\over(m^*-m_N)
\omega}e^{-k^2/6\omega^2}f_{\pi NN}\simeq -1.71 f_{\pi NN}$\\ \hline
 \\
$f_{\pi N\Delta}^{(1620)}=-{\sqrt{3}\over 15}{\omega\over
(m^*-m_N)}
({k^2\over \omega^2}+{9\omega\over 2 m_q})
e^{-k^2/6\omega^2}f_{\pi NN}\simeq -0.34 f_{\pi NN}$\\
 \\
$f_{\pi N\Delta}^{(1700)}={12\over 5}{m^* m_N\over
(m^*-m_N)\omega}e^{-k^2/6\omega^2}f_{\pi NN}\simeq 2.6 f_{\pi NN}$\\ \hline
 \\
$f_{\pi NN^*}^{(1650)}={\sqrt{2}\over 15}{\omega\over
(m^*-m_N)}
({k^2\over \omega^2}+{9\omega\over 2 m_q})
e^{-k^2/6\omega^2}f_{\pi NN}\simeq 0.28 f_{\pi NN}$\\
 \\
$f_{\pi NN^*}^{(1700)}={4\sqrt{15}\over 75}{m^* m_N\over
(m^*-m_N)\omega}e^{-k^2/6\omega^2}f_{\pi NN}\simeq 0.22 f_{\pi NN}$\\
 \\
$f_{\pi NN^*}^{(1675)}={10\over 25}{m_\pi\over \omega}
e^{-k^2/6\omega^2}f_{\pi NN}\simeq 0.09 f_{\pi NN}$\\ \hline
 \\
$f_{\pi NN^*}^{(1720)}={\sqrt{10}\over 30}
({k^2\over \omega^2}+{15\omega_\pi \over 2 m_q})
e^{-k^2/6\omega^2}f_{\pi
NN}\simeq -1.05$\\
 \\
$f_{\pi NN^*}^{(1680)}=-{\sqrt{2}\over 6}{m^*\over (m^*+m_N)}({m_N\over
m_\pi})({m_\pi\over \omega})^2e^{-k^2/6\omega^2}f_{\pi NN}\simeq -0.12$\\
\hline 
\end{tabular}

\begin{tabular}{|l|c|l|} \hline
\multicolumn{3}{|l|}{Table 5. Transition matrix elements of the
isoscalar quark charge $\sum_{q}e^{-i\vec k\cdot \vec r_q}$ and}\\
\multicolumn{3}{|l|}{transverse spin current $\sum_{q}
\vec\sigma^q\cdot
(\vec k\times \vec \epsilon)e^{-i\vec
k\cdot \vec r_q}$ operators between the proton and nucleon}\\
\multicolumn{3}{|l|}{resonances
for charge states $+1$ with $s_z=+1/2$.}\\ \hline
 & & \\
 transition & charge & spin current\\ \hline
 & & \\
$<p,{1\over 2}|O|p,{1\over 2}>$ & 3 & $(\vec k\times \vec
\epsilon)_3$\\ \hline
 & & \\
$<p,{1\over 2}|O|N(1440)^+,{1\over 2}>$ & ${\sqrt{3}\over 6}
{\vec k^2\over
\omega^2}e^{-k^2/6\omega^2}$ & ${\sqrt{3}\over 18}{k^2\over
\omega^2}e^{-k^2/6\omega^2}(\vec k\times \vec \epsilon)_3$\\ \hline
 & & \\
$<p,{1\over 2}|O|N(1535)^+,{1\over 2}>$ & 0 & $-{\sqrt{2}\over 3}{
\vec k^2\over
\omega}e^{-k^2/6\omega^2}\epsilon_3$ \\
 & & \\
$<p,{1\over 2}|O|N(1520)^+,{1\over 2}>$ & 0 & $i{2\over 3}{1\over
\omega}e^{-k^2/6\omega^2}[\vec k_3(\vec k\times \vec
\epsilon)_3+k_+(\vec k\times \vec \epsilon)_-]$ \\ \hline
 & & \\
$<p,{1\over 2}|O|N(1650)^+,{1\over 2}>$ & 0 & ${\sqrt{2}\over
6}{\vec k^2\over \omega}e^{-k^2/6\omega^2}\epsilon_3$ \\
 & & \\
$<p,{1\over 2}|O|N(1700)^+,{1\over 2}>$ & 0 & $i{\sqrt{10}\over 15}
{1\over\omega}e^{-k^2/6\omega^2}[k_3(\vec k\times \vec
\epsilon)_3+k_+(\vec k\times \vec \epsilon)_-$ \\
& & \\
 & & $+3i\epsilon_3\vec k^2]$ \\
& & \\
$<p,{1\over 2}|O|N(1675)^+,{1\over 2}>$ & 0 & $-i{3\sqrt{10}\over
10}{k_3\over \omega}e^{-k^2/6\omega^2}(\vec k\times \vec
\epsilon)_3$\\ \hline
&&\\
$<p,{1\over 2}|O|N(1720)^+,{1\over 2}>$ & $0
$ & $-{\sqrt{30}\over 180}{k^2\over
\omega^2}e^{-k^2/6\omega^2}(\vec k\times \vec \epsilon)_3$\\
&&\\
$<p,{1\over 2}|O|N(1680)^+,{1\over 2}>$ & $
-{\sqrt{10}\over 20}{3k_3^2-\vec k^2\over \omega^2}
e^{-k^2/6\omega^2}
$ & $-i{\sqrt{30}\over 45}{k^3\over
\omega^2}e^{-k^2/6\omega^2}$\\
& & \\
& & $\sum_q(3,1,q,-q\vert 3,0)\sqrt{4\pi\over 7}Y_{3q}(\hat k)
\vec \epsilon_{-q}$ \\ \hline
\end{tabular}

\begin{tabular}{|l|c|l|} \hline
\multicolumn{3}{|l|}{Table 6. Transition matrix elements of the
isovector quark charge $\sum_{q}\tau_3^qe^{-i\vec k\cdot \vec r_q}$}\\
\multicolumn{3}{|l|}{and transverse current operator
$\sum_{q}\tau_3^q\vec \sigma^q\cdot (\vec k\times \vec
\epsilon)e^{-i\vec k\cdot \vec r_q}$ between the proton and}\\
\multicolumn{3}{|l|}{nucleon resonances
with charge state $+1$ with $s_z=+1/2$.}\\ \hline
 & & \\
transition & \phantom{$\quad\quad$}charge\phantom{$\quad\quad$}
& spin current\\ \hline
 & & \\
$<p,{1\over 2}|O|p,{1\over 2}>$ & 1 & ${5\over 3}
(\vec k\times \vec \epsilon)_3$\\ 
 & & \\
$<p,{1\over 2}|O|\Delta(1232)^+,{1\over 2}>$ & 0 & ${4\sqrt{2}\over
3}e^{-k^2/6\omega^2}(\vec k\times
\vec \epsilon)_3$\\ \hline
&&\\
$<p,{1\over 2}|O|N(1440)^+,{1\over 2}>$ & ${\sqrt{3}\over 18}{k^2\over
\omega^2}e^{-k^2/6\omega^2}$ & ${5\sqrt{3}\over
54}{\vec k^2\over \omega^2}e^{-k^2/6\omega^2}
(\vec k\times \vec \epsilon)_3$\\
 & & \\
$<p,{1\over 2}|O|\Delta(1600)^+,{1\over 2}>$ & 0 & ${3\sqrt{6}\over
27}{k^2\over \omega^2}e^{-\vec k^2/6\omega^2}(\vec k\times
\vec \epsilon)_3$\\ \hline
&&\\
$<p,{1\over 2}|O|N(1535)^+,{1\over 2}>$ & $-i{\sqrt{2}\over 3}{k_3\over
\omega}e^{-k^2/6\omega^2}$ & $-{2\sqrt{2}\over 9}{\vec k^2\over
\omega}e^{-k^2/6\omega^2}\epsilon_3$\\
 & & \\
$<p,{1\over 2}|O|N(1520)^+,{1\over 2}>$ & 0 & $i{4\over 9}{1\over
\omega}e^{-k^2/6\omega^2}[k_3(\vec k\times \vec \epsilon)_3+k_+(\vec
k\times \vec \epsilon)_-]$\\ \hline
&&\\
$<p,{1\over 2}|O|\Delta(1620)^+,{1\over 2}>$ & $i{\sqrt{2}\over
3}{k_3\over \omega}e^{-k^2/6\omega^2}$& ${\sqrt{2}\over 18}{\vec k^2\over
\omega}e^{-k^2/6\omega^2}\epsilon_3$\\
 & & \\
$<p,{1\over 2}|O|\Delta(1700)^+,{1\over 2}>$ & 0 & ${i\over 9}{1\over
\omega}e^{-k^2/6\omega^2}[k_3(\vec k\times \vec \epsilon)_3+k_+(\vec
k\times \vec \epsilon)_-]$ \\ \hline
&&\\
$<p,{1\over 2}|O|N(1650)^+,{1\over 2}>$ & $-{i\over 6}
{k_3\over \omega}e^{-k^2/6\omega^2}$& $-{\sqrt{2}\over 18}{\vec k^2\over
\omega}e^{-k^2/6\omega^2}\epsilon_3$\\
 & & \\
$<p,{1\over 2}|O|N(1700)^+,{1\over 2}>$ & 0 & $-i{\sqrt{10}\over
45}{1\over \omega}e^{-k^2/6\omega^2}[k_3(\vec k\times \vec
\epsilon)_3+k_+(\vec k\times \vec \epsilon)_-$\\
& &$+3i\epsilon_3\vec k^2]$\\
& & \\
$<p,{1\over 2}|O|N(1675)^+,{1\over 2}>$ & 0 & $i{\sqrt{10}\over
10}{g_{\rho NN^*}^{(1675)}\over m_\rho^2}k_3
(\vec k\times \vec \epsilon)_3$
\\ \hline
&&\\
$<p,{1\over 2}|O|N(1720)^+,{1\over 2}>$ & $0
$ & $-{\sqrt{30}\over 108}{k^2\over
\omega^2}e^{-k^2/6\omega^2}(\vec k\times \vec \epsilon)_3$\\
&&\\
$<p,{1\over 2}|O|N(1680)^+,{1\over 2}>$ & $
-{\sqrt{10}\over 60}{3k_3^2-\vec k^2\over \omega^2}
e^{-k^2/6\omega^2}
$ & $-i{\sqrt{30}\over 27}{k^3\over
\omega^2}e^{-k^2/6\omega^2}$\\
& & $\sum_q(3,1,q,-q\vert 3,0)\sqrt{4\pi\over 7}Y_{3q}(\hat k)
\vec \epsilon_{-q}$ \\ \hline
\end{tabular}

\begin{tabular}{|l|c|l|} \hline
\multicolumn{3}{|l|}{Table 7. Transition matrix elements $<p,{1\over
2}|{\cal L}_{\omega NN^*}|N^{*+},{1\over 2}>$ of the $\omega$-meson}\\
\multicolumn{3}{|l|}{transition couplings in eqs. (3.6). 
Here $\mu$ is defined as}\\
\multicolumn{3}{|l|}{ $\mu=2m_Nm^*/(m_N+m^*)$.}\\ \hline
 & & \\
transition & charge & spin current \\ \hline
 & & \\
$<p,{1\over 2}|{\cal L}_{\omega NN^*}^{(1440)}|N(1440)^+,{1\over 2}>$ &
${\vec k^2\over m_\omega^2}\,
g_{\omega NN^*}^{(1440)}$ 
& $-i{1\over  2\mu}{\vec k^2\over m_\omega ^2}
(\vec k\times\vec\epsilon)_3\,
g_{\omega NN^*}^{(1440)}
$ \\
&&\\
 \hline
 & & \\
$<p,{1\over 2}|{\cal L}_{\omega NN^*}^{(1535)}|N(1535)^+,{1\over 2}>$ &
0 & ${\vec k^2\over m_\omega^2}\epsilon_3\, g_{\omega NN^*}^{(1535)}$\\
 & & \\
$<p,{1\over 2}|{\cal L}_{\omega NN^*}^{(1520)}|N(1520)^+,{1\over 2}>$ &
$0$ &
$i{\sqrt{6}\over 3 m_\omega^2}[k_3(\vec
k\times \vec \epsilon)_3+k_+(\vec k\times \vec \epsilon)_-]\,
g_{\omega NN^*}^{(1520)}$\\
&&\\
 \hline
 & & \\
$<p,{1\over 2}|{\cal L}_{\omega NN^*}^{(1650)}|N(1650)^+,{1\over 2}>$ &
0 & ${\vec k^2\over m_\omega^2}\epsilon_3\,
g_{\omega NN^*}^{(1650)}
$\\
 & & \\
$<p,{1\over 2}|{\cal L}_{\omega NN^*}^{(1700)}|N(1700)^+,{1\over 2}>$ &
$0$ &
$i{\sqrt{6}\over 3 m_\omega^2}[k_3(\vec
k\times \vec \epsilon)_3+k_+(\vec k\times \vec \epsilon)_-]\,
g_{\omega NN^*}^{(1700)}$\\
 & & \\
$<p,{1\over 2}|{\cal L}_{\omega NN^*}^{(1675)}|N(1675)^+,{1\over 2}>$ &
0 & $-i{3\sqrt{10}\over 10}{1\over
m_\omega^2}k_3(\vec k\times \vec \epsilon)_3\, g_{\omega NN^*}^{(1675)}$ \\
&&\\
 \hline
&&\\
$<p,{1\over 2}|{\cal L}_{\omega NN^*}^{(1720)}|N(1720)^+,{1\over 2}>$ &
$0$ 
& $i{1\over  2\mu}\sqrt{2\over 3}{\vec k^2\over m_\omega ^2}
(\vec k\times\vec\epsilon)_3\,
g_{\omega NN^*}^{(1720)}
$ \\
&&\\
$<p,{1\over 2}|{\cal L}_{\omega NN^*}^{(1680)}|N(1680)^+,{1\over 2}>$ &
${\sqrt{10}\over 10}{3k_3^2-\vec k^2\over m_\omega^2}
g_{\omega NN^*}^{(1680)}$ 
& $-{1\over \mu}{\sqrt{30}\over 15}{\vec k^3\over m_\omega ^2}
g_{\omega NN^*}^{(1680)}$ \\
& & \\
& & $(3,1,q,-q\vert 3,0)\sqrt{{4\pi\over 7}}Y_{3 q}(\hat k)
\epsilon_{-q}$\\
& & \\ \hline
\end{tabular}

\begin{tabular}{|l|} \hline
Table 8. The $\omega$-meson transition coupling constants in terms\\ 
of the $\omega NN$ coupling constant $g_{\omega NN}$. The numerical
values  \\
correspond to $\vert \vec k\vert =0$. \\ \hline
 \\
$g_{\omega NN^*}^{(1440)}={\sqrt{3}\over 18}
{m_\omega ^2\over \omega^2}
e^{-\vec k^2/6\omega^2}g_{\omega NN}\simeq 5.5$ \\
\\ \hline
 \\
$g_{\omega NN^*}^{(1535)}=-{\sqrt{2}\over 18}{m_\omega^2
\over \omega m_q}e^{-\vec k^2/6\omega^2}g_{\omega NN}
\simeq -4.5$ \\
 \\
$g_{\omega NN^*}^{(1520)}={\sqrt{6}\over 18}{m_\omega^2\over \omega
m_q}e^{-\vec k^2/6\omega^2}g_{\omega NN} \simeq 7.7$ \\
\\ \hline
 \\
$g_{\omega NN^*}^{(1650)}={\sqrt{2}\over 36}{m_\omega^2
\over \omega m_q}e^{-\vec k^2/6\omega^2} g_{\omega NN}
\simeq 2.2$ \\
 \\
$g_{\omega NN^*}^{(1700)}={\sqrt{15}\over 90}
{m_\omega^2\over \omega m_q}e^{-\vec k^2/6\omega^2}
g_{\omega NN}\simeq 2.4$ \\
 \\
$g_{\omega NN^*}^{(1675)}={1\over 6}
{m_\omega^2\over \omega m_q}
e^{-\vec k^2/6\omega^2}g_{\omega NN}\simeq 9.4$ \\
\\ \hline
\\
$g_{\omega NN^*}^{(1720)}=-{\sqrt{5}\over 180}
{\mu\over m_q}{m_\omega^2\over \omega^2 }
e^{-\vec k^2/6\omega^2}g_{\omega NN}\simeq -2.7$ \\ 
\\
$g_{\omega NN^*}^{(1680)}=-{1\over 18}
{\mu\over m_q}{m_\omega^2\over \omega^2 }
e^{-\vec k^2/6\omega^2}g_{\omega NN}\simeq -12.2$ \\ 
\\ \hline
\end{tabular}

\begin{tabular}{|l|c|l|} \hline
\multicolumn{3}{|l|}{Table 9. Transition matrix elements $<p,{1\over
2}|{\cal L}_{\rho NN^*}|N^{*+},{1\over 2}>$ of the $\rho$-meson}\\
\multicolumn{3}{|l|}{transition couplings in eqs. (3.7)
Here $\mu$ is defined as}\\ 
\multicolumn{3}{|l|}{$\mu=2m^*m_N/(m^*+m_N)$.
}\\ \hline
 & & \\
transition & charge & spin current\\ \hline
 & & \\
$<p,{1\over 2}|{\cal L}_{\rho N\Delta}|\Delta (1232)^+,{1\over 2}>$ & 0
& $i{1\over 3\mu}(\vec k\times \vec \epsilon)_3\,
g_{\rho N\Delta}
$ \\ \hline
 & & \\
$<p,{1\over 2}|{\cal L}_{\rho NN^*}^{(1440)}|N(1440)^+,{1\over 2}>$ &
${\vec k^2\over m_\rho ^2}\,
g_{\rho NN^*}^{(1440)}$ 
& $-i{1 \over 2\mu}
{\vec k^2\over m_\rho ^2}(\vec k\times \vec \epsilon)_3\,
g_{\rho NN^*}^{(1440)}(1+\kappa_{\rho NN} 
^{(1440)})
$ \\
& & \\
$<p,{1\over 2}|{\cal L}_{\rho N\Delta^*}^{(1600)}|\Delta(1600)^+,{1\over
2}>$ & 0 & $i{1 \over 3\mu}
{\vec k^2\over m_\rho^2}(\vec k\times \vec
\epsilon)_3\,g_{\rho N\Delta^*}^{(1600)}
$ \\ \hline
 & & \\
$<p,{1\over 2}|{\cal L}_{\rho NN^*}^{(1535)}|N(1535)^+,{1\over 2}>$ &
$0$ & $
{\vec k^2\over m_\rho^2}\epsilon_3\,g_{\rho NN^*}^{(1535)}$\\
 & & \\
$<p,{1\over 2}|{\cal L}_{\rho NN^*}^{(1520)}|N(1520)^+,{1\over 2}>$ &
$0$ &
$i{\sqrt{6}\over 3 m_\rho^2}[k_3(\vec
k\times \vec \epsilon)_3+k_+(\vec
k\times \vec \epsilon)_-]\,g_{\rho NN^*}^{(1520)}$
\\ & & \\ \hline
&&\\
$<p,{1\over 2}|{\cal L}_{\rho N\Delta^*}^{(1620)}|\Delta(1620)^+,{1\over
2}>$ & $0$ &
$\sqrt{{2\over 3}}{\vec k^2\over m_\rho^2 }
g_{\rho N\Delta^*}^{(1620)}\epsilon_3$\\
 & & \\
$<p,{1\over 2}|{\cal L}_{\rho N\Delta^*}^{(1700)}|\Delta(1700)^+,{1\over
2}>$ & $0$ &
$i{2\over 3 m_\rho^2}[k_3(\vec k\times \vec \epsilon)_3+k_+
(\vec k\times \vec \epsilon)_-]\,g_{\rho N\Delta^*}^{(1700)}$\\
&&\\
 \hline
&&\\
$<p,{1\over 2}|{\cal L}_{\rho NN^*}^{(1650)}|N(1650)^+,{1\over 2}>$
& $0$ & $
{\vec k^2\over m_\rho^2}\epsilon_3\,
g_{\rho N\Delta^*}^{(1650)}
$\\
 & & \\
$<p,{1\over 2}|{\cal L}_{\rho NN^*}^{(1700)}|N(1700)^+,{1\over 2}>$ &
$0$ &
$i{\sqrt{6}\over 3m_\rho^2}[
k\times \vec \epsilon)_3+k_+(\vec k\times
\vec \epsilon)_-]\,g_{\rho NN^*}^{(1700)}$\\
&&\\
$<p,{1\over 2}|{\cal L}_{\rho NN^*}^{(1675)}|N(1675)^+,{1\over 2}>$ &0 &
$-i{3\sqrt{10}\over 10}{1\over m_\rho^2}
k_3(\vec k\times\epsilon)_3\, g_{\rho NN^*}^{(1675)}$ \\ \hline
&&\\
$<p,{1\over 2}|{\cal L}_{\rho NN^*}^{(1720)}|N(1720)^+,{1\over 2}>$ &
$0$ 
& $i{1\over  2\mu}\sqrt{2\over 3}{\vec k^2\over m_\rho ^2}
(\vec k\times\vec\epsilon)_3\,
g_{\rho NN^*}^{(1720)}
$ \\
&&\\
$<p,{1\over 2}|{\cal L}_{\rho NN^*}^{(1680)}|N(1680)^+,{1\over 2}>$ &
${\sqrt{10}\over 10}{3k_3^2-\vec k^2\over m_\rho^2}
g_{\rho NN^*}^{(1680)}$ 
& $-{1\over \mu}{\sqrt{30}\over 15}{\vec k^3\over m_\rho ^2}
g_{\rho NN^*}^{(1680)}$ \\
& & \\
& & $(3,1,q,-q\vert 3,0)\sqrt{{4\pi\over 7}}Y_{3 q}(\hat k)
\epsilon_{-q}$
\\ \hline
\end{tabular}

\begin{tabular}{|l|} \hline
Table 10. The $\rho$-meson transition coupling constant in terms of the\\
$\rho NN$ coupling constants $g_{\rho NN}$.
The numerical values correspond\\
to $\vert \vec k\vert=0$.
\\ \hline
\\
$g_{\rho N\Delta}={6\sqrt{2}\over 5}{\mu\over m_q}
e^{-\vec k^2/6 \omega^2}g_{\rho NN}\simeq 8.7$ \\ \hline
\\
$g_{\rho NN^*}^{(1440)}={\sqrt{3}\over 18} 
{m_\rho^2\over \omega^2}
e^{-\vec k^2/6\omega^2}g_{\rho NN}\simeq 
1.76,\,\,\,\, \kappa_{\rho NN^*}
^{1440}=4$ \\
\\
$g_{\rho N\Delta^*}^{(1600)}={\sqrt{6}\over 10}{\mu\over
m_q}{m_\rho^2\over \omega^2}
e^{-\vec k^2/6\omega^2}g_{\rho NN}\simeq 17.0$ \\ \hline
\\
$g_{\rho NN^*}^{(1535)}=-{\sqrt{2}\over 9}
{m_\rho^2\over \omega m_q}e^{-\vec k^2/6\omega^2}g_{\rho NN}
\simeq -2.9$ \\ 

\\
$g_{\rho NN^*}^{(1520)}={\sqrt{6}\over 9}{m_\rho^2\over\omega m_q}
e^{-\vec k^2/6\omega^2}g_{\rho NN^*}\simeq 4.5 $ \\ \hline
\\
$g_{\rho N\Delta^*}^{(1620)}={\sqrt{3}\over 36}{m_\rho^2\over
\omega m_q}e^{-\vec k^2/6\omega^2} g_{\rho NN}\simeq 0.88 
$ \\
\\
$g_{\rho N\Delta^*}^{(1700)}={1\over 12}{m_\rho^2\over \omega m_q}
e^{-\vec k^2/6\omega^2}g_{\rho N N}\simeq 1.5$ \\ \hline
\\
$g_{\rho NN^*}^{(1650)}=-{\sqrt{2}\over 36}{m_\rho^2\over 
\omega m_q}e^{-\vec k^2/6\omega^2}g_{\rho NN}\simeq -0.72$ \\
\\
$g_{\rho NN^*}^{(1700)}=-{\sqrt{5}\over 90}{m_\rho^2\over
\omega m_q}e^{-\vec k^2/6\omega^2}g_{\rho NN}\simeq -0.45$  \\
\\
$g_{\rho NN^*}^{(1675)}=-{1\over 6}
{m_\rho^2\over \omega m_q}e^{-\vec k^2/6\omega^2}
g_{\rho NN}\simeq -3.0$  \\ \hline
\\
$g_{\rho NN^*}^{(1720)}=-{\sqrt{5}\over 36}
{\mu\over m_q}{m_\rho^2\over \omega^2 }
e^{-\vec k^2/6\omega^2}g_{\rho NN}\simeq -4.4$ \\ 
\\
$g_{\rho NN^*}^{(1680)}=-{5\over 18}
{\mu\over m_q}{m_\rho^2\over \omega^2 }
e^{-\vec k^2/6\omega^2}g_{\rho NN}\simeq -19.6$ \\ 
\\ \hline
\end{tabular}

\end{document}